\documentclass[amsmat,amssymb,amsfonts,aps,prl,twocolumn,showpacs,superscriptaddress]{revtex4-1}
\usepackage{graphicx}
\usepackage{dcolumn}
\usepackage{bm}
\newcommand{\beq}{\begin{equation}}  
\newcommand{\eeq}{\end{equation}}  
\newcommand{\beqa}{\begin{eqnarray}}  
\newcommand{\eeqa}{\end{eqnarray}}  
  
\usepackage{xcolor}

\begin{document}

\title{Local probe of fractional edge states of  $S=1$ Heisenberg spin chains }


\author{F. Delgado}
\affiliation{International Iberian Nanotechnology Laboratory (INL),
Av. Mestre Jos\'e Veiga, 4715-310 Braga, Portugal}

\author{C. D. Batista}
\affiliation{Theoretical Division, T-4 and CNLS, Los Alamos National Laboratory, Los Alamos, New Mexico 87545, USA}

\author{J. Fern\'andez-Rossier\footnote{Permanent address: Departamento de F\'{i}sica Aplicada, Universidad de Alicante,  Spain}}
\affiliation{International Iberian Nanotechnology Laboratory (INL),
Av. Mestre Jos\'e Veiga, 4715-310 Braga, Portugal}


\begin{abstract}
 Spin chains are among the simplest physical systems in which electron-electron interactions induce novel states of matter. Here we show that the combination of atomic scale engineering and spectroscopic capabilities of 
state of the art scanning tunnel microscopy enables probing the fractionalized edge states of individual atomic scale $S=1$ spin chains. These edge states arise from the  topological order of the ground state in the Haldane phase. We also show that the Haldane gap  and  the spin-spin  correlation length can be measured with the same technique.
 \end{abstract}

\date{\today}

 \maketitle



The basic electronic properties of  large classes of materials, such as good metals and semiconductors, 
can be described with an independent electron picture because the Coulomb electron-electron interaction plays a marginal role. In contrast, the single-electron picture is not applicable to strongly correlated materials, such as Mott insulators and their descendants, which exhibit  a variety of fascinating phenomena including high temperature superconductivity and colossal magnetoresistance~\cite{Dagotto_Hotta_pr_2001,Saxena_book_2012}. 
Modeling the electronic properties of  these strongly correlated materials can be notoriously difficult. 
However,  different analytical and numerical tools, developed for this purpose, provide a great deal of insight for strongly correlated one-dimensional systems. The study of quantum spin chains played a central role in this context. Exact analytical solutions~\cite{Bethe_zur_1931} and  efficient numerical methods~\cite{White_prl_1992} revealed a plethora of collective phenomena that triggered new paradigms,  
such as quantum spin liquids \cite{Haldane_prl_1983,Anderson_science_1987},  fractionalized spin excitations~\cite{Haldane_prl_1991,Wen_intj_1992,Wen_advphys_1995} and hidden topological order\cite{Afflect_Haldane_prb_1987,Wen_book_2004}. The gapped Haldane phase of integer spin antiferromagnetic (AFM) Heisenberg chains is a remarkable example of a novel quantum state induced by electron-electron interactions~\cite{Haldane_prl_1983}.  

In analogy with the in-gap edge states  of  topologically ordered phases \cite{Wen_intj_1992,Wen_advphys_1995},   the $S=1$  AFM Heisenberg chain exhibits fractional $S=1/2$ edge states, whose energy  lies inside the Haldane gap. These in-gap states  behave like  two $S=1/2$ spins that are  localized at each end of the open chain~\cite{Hagiwara_Katsumata_prl_1990,Glraum_Geschwind_prl_1991,White_prb_1996},  in spite of the fact that the elementary building blocks of the model are $S=1$ spins.
  
The existence of these $S=1/2$ edge states was revealed by  Electron Spin Resonance (ESR)
\cite{Hagiwara_Katsumata_prl_1990,Glraum_Geschwind_prl_1991,Yoshida_Shiraki_prl_2005}
and neutron scattering 
\cite{Buyers_Morra_prl_1986} 
experiments performed on quasi-one-dimensional $S=1$ materials.  Measurements of thermodynamic properties  
properties are also consistent with the existence of these end-sates \cite{Ramirez_Cheong_prl_1994, Batista_Hallberg_prb_1998,Shapira_Bindilatti_jap_2002}. However, these experiments can only probe  a thermodynamically large number of open $S=1$ chains of different lengths, which arise from chemical substitution of the $S=1$ ion by a non-magnetic ion.   

\begin{figure}
\includegraphics[width=1.\linewidth,angle=0]{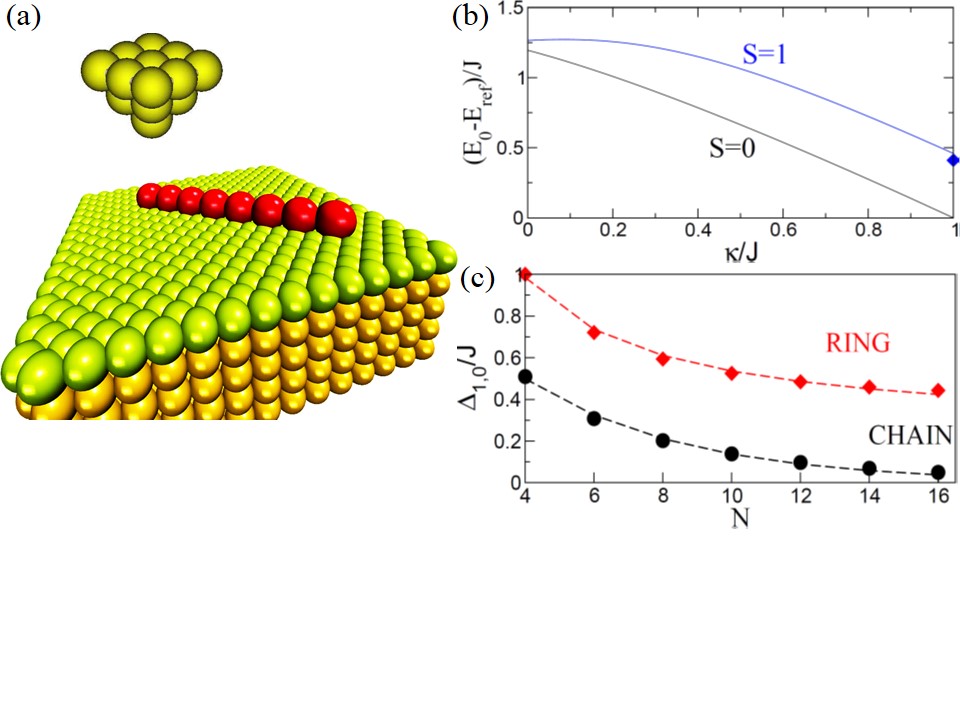}
\caption{ (Color online)\label{fig1} (a) Scheme of an STM tip probing the spin excitation of a chain of magnetic adatoms deposited on top an insulating layer as in Ref.~\cite{Hirjibehedin_Lutz_Science_2006}. (b) Adiabatic evolution of the low energy spectra for an AFM Heisenberg chain of  $N=14$ $S=1$ spins from PBC to OBC. The lowest excitation energy for the infinite ring is equal to the Haldane gap (diamond), while the open chain becomes gapless for $N \to \infty$ because of the $S=1/2$ states that are localized on each edge.  (c) Size dependence of the singlet-triplet energy gap for PBC and OBC in absence of anisotropy (only even sizes are plotted).}
\end{figure}

In this Letter we propose a radically different approach based on the recent progress in  atomic scale manipulation 
and single atom inelastic electron tunneling spectroscopy (IETS) of spin excitations 
the can be measured with a scanning tunneling microscope (STM).  The STM   can be used as a tool for arranging the magnetic ions in a single linear chain~\cite{Hirjibehedin_Lutz_Science_2006,Serrate_Ferriani_natnano_2010,Khajetoorians_Wiebe_science_2011,Khajetoorians_Wiebe_natphys_2012,Loth_Baumann_science_2012}. The spectral features of the linear chain remain sharp  if  the magnetic adatoms are deposited on a thin 
insulating layer, like a single monolayer of Cu$_2$N coating a Cu(100)  substrate  ~\cite{Hirjibehedin_Lutz_Science_2006,Hirjibehedin_Lin_Science_2007,Loth_Baumann_science_2012},
that reduces the exchange coupling to the underlying metallic substrate.  Under these conditions, it is possible to perform atomically resolved  spin IETS~\cite{Hirjibehedin_Lutz_Science_2006} [see Fig.~\ref{fig1}(a)]. Indeed, STM-IETS measurements of  Mn, Fe or Co atoms deposited on a thin insulating layer revealed quantized spin values of $S=5/2, 2$ and $3/2$, respectively,  that are consistent with Hund's rules \cite{Hirjibehedin_Lin_Science_2007}. 
The same technique has revealed  spin excitations of chains containing  up to 10  $S=5/2$ Mn atoms, which are are well described by an AFM Heisenberg model with single-ion anisotropy terms \cite{Hirjibehedin_Lutz_Science_2006,  Rossier_prl_2009}.  In addition,  intensity modulations of the  IETS-STM signal on different atoms of the chain have been theoretically predicted~\cite{Rossier_prl_2009} and experimentally observed \cite{Otte_PhD_2008,Bryant_Spinelly_arXiv_2013}.

Here we demonstrate that IETS  performed with an STM on a family of $S=1$ spin chains of 
increasing lengths $N \leq 16$,  which are well within the experimental range~\cite{Hirjibehedin_Lutz_Science_2006,Loth_Baumann_science_2012},  provides a  local probe for the $S=1/2$ edge states of the Haldane phase.  
We also show that our results are  robust against the unavoidable presence  of  single-ion anisotropy.  Our  proposal suggests that atomically engineered spin chains can be used as ``quantum simulators'' for probing exotic correlated states of matter, such as  gapped quantum spin liquids. 
  
We start by considering  the   Heisenberg Hamiltonian for  a  chain (ring) of $N$ anisotropic $S=1$ spins 
\begin{equation}
H=\sum_{i=1}^N h_0(i) + J \sum_{i=1}^{N-1} {\bf S}_i\cdot {\bf S}_{i+1}+\kappa {\bf S}_N\cdot {\bf S}_1,
\label{hamilT}
\end{equation}
where
\begin{equation}
h_0(i)= D (S^z_i)^2 + E\left[ (S^x_i)^2-(S^y_i)^2\right]
\end{equation}
is the single-ion anisotropy contribution to $H$, with axial anisotropy $D$ and transverse anisotropy $E$.
$J,\kappa\ge0$ are the  antiferromagnetic   exchange  inter-adatom energies. 
The two relevant limiting cases correspond to open boundary conditions (OBC) for $\kappa=0$   and periodic boundary conditions (PBC) for $\kappa=J$. 
 We shall denote the eigenvectors of this Hamiltonian as $|m\rangle$
and  the eigenvalues as $E_m$ in increasing order, $E_{m-1} \leq E_m$, with $m=0$ corresponding to the ground state. The low-energy eigenvalues and eigenstates are obtained by direct numerical diagonalization 
of  $H$ for $N \leq 16$. 

We first discuss the results for the isotropic case $D=E=0$. This is a good starting point in the case of $S=5/2$ Mn chains because the orbital moment is quenched for the 5 polarized $3d$ electrons~\cite{Hirjibehedin_Lutz_Science_2006, Rossier_prl_2009}.  
Figure~\ref{fig1}(b)  shows  the low energy spectra of $H$ as a function of $\kappa/J$ for $N=14$. The ground state is a singlet in both cases, as expected for even $N$, while the first excited state has total spin $S=1$. However, the energy gap between them decreases from a value that is close to the Haldane gap of the ring ($\kappa=1$) to a much smaller value in the open chain limit. The finite size scaling of this gap $\Delta_{1,0}=E_1-E_0$  sheds light on the nature of the lowest energy excited states. For the ring, the size dependence of the energy gap is well fitted by the expression $\Delta_{1,0}=\Delta_H + \eta J/N^2$ 
($\Delta_H=0.411J$ is the Haldane gap marked by a diamond in Fig.~\ref{fig1}(b) and 
 $\eta\approx 9.84$), as expected from the quadratic dispersion relation of the single-magnon excited states~\cite{White_Huse_prb_1993}.
The finite size scaling of the singlet-triplet gap  is qualitatively different for open chains:
\beq
\Delta_{1,0}=\Delta_0 e^{-N/\xi},
\label{xi}
\eeq
where  $\xi\approx 4.65$ is the spin-spin correlation length and $\Delta_0=1.2 J$  in perfect agreement with previous results~\cite{Kennedy_jphys_1999,Sakai_Takahashi_prb_1991,Golinelli_Jolicoeur_prb_1992}.  For a characteristic exchange constant
of  $6$meV (the case of Mn chains),  the excitation energy for a chain of $N=16$ spins would be $\Delta_{1,0}\simeq$ 230 $\mu$eV. Excitations in this energy scale have been resolved previously with STM-IETS~\cite{Hirjibehedin_Lin_Science_2007}.

\begin{figure}
\includegraphics[width=1.\linewidth,angle=0]{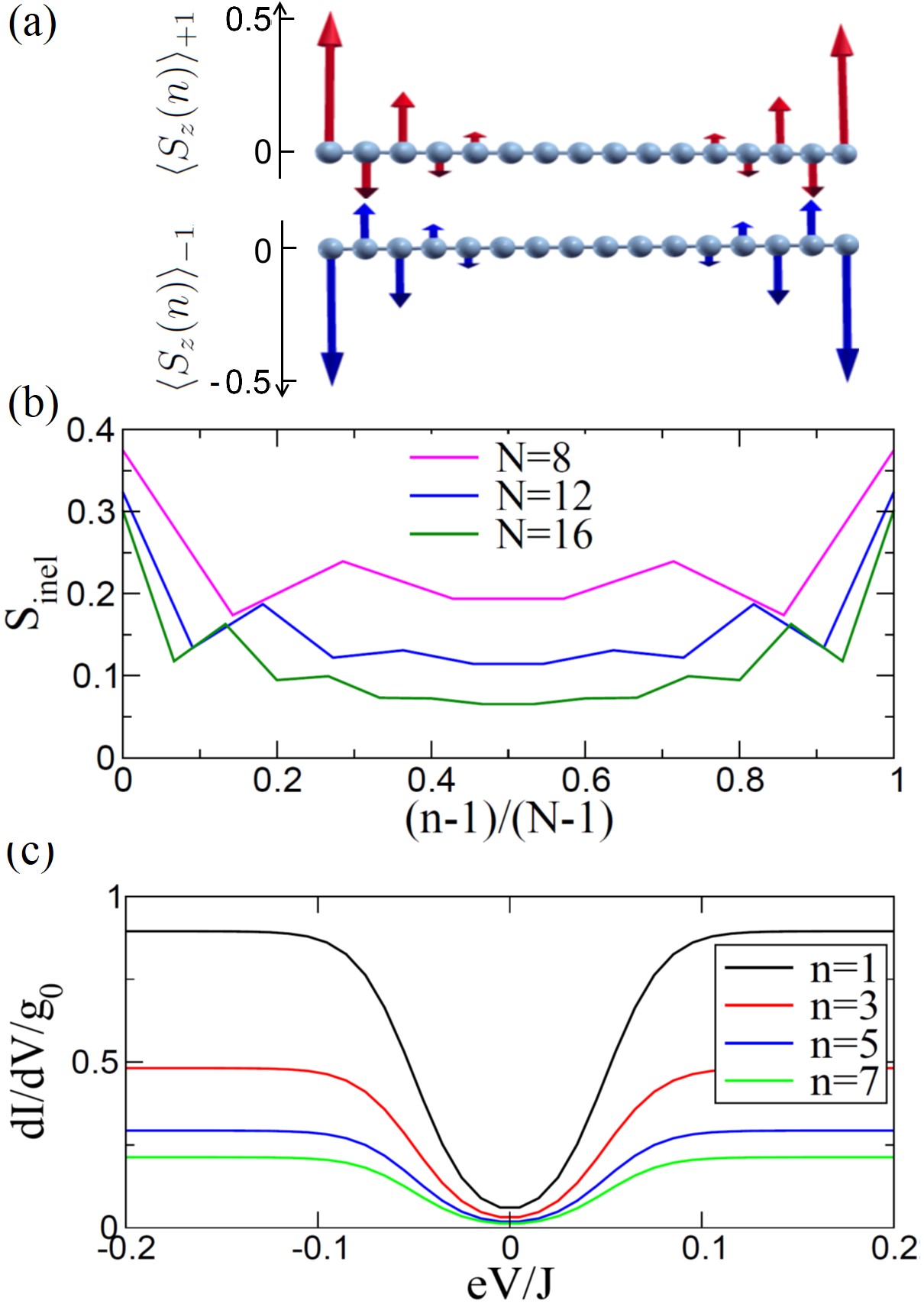}
\caption{ (Color online) (a) Expectation value $\langle S^z_i \rangle_m$ for the two triplet states with $S_z=\pm 1$.
The spin density is peaked at the edges.    
(b) Inelastic signal $S_{\rm inel}=\sum_{a}\sum_{m=1}^3|\langle 0|S^a_n|m\rangle|^2$ versus the normalized atomic position $(n-1)/(N-1)$ for different chain sizes. $D=E=0$ in all cases.
(c) IETS on atoms 1, 3, 5 and 7 for the  $N=16$ chain at $T=0.01J$.  The step of $dI/dV$ is at  $eV=\Delta_{1,0}$.  The height of the step
is controlled by   $S_{\rm inel} $  shown in panel (b).}
\label{fig2}
\end{figure}

The exponential decay of Eq. (\ref{xi})  is known to  arise~\cite{Batista_Hallberg_prb_1998} from the coupling of two  $S=1/2$ modes localized at the edges.  The gap $\Delta_{1,0}$ corresponds to the effective  coupling between both edge modes and the existence of these modes is illustrated by the site dependence of  $\langle 1, \pm 1 |S^z_i| 1,\pm 1\rangle$  shown in  Fig.~\ref{fig2}(a).  The expectation value  is $\pm \frac{1}{2}$ for the boundary atoms  and it decays exponentially towards the center of the chain with $\pi$ oscillations that reflect the AFM nature of the exchange coupling \cite{Yamamoto94}.    

Because the mean value of the  magnetization  on each site is not an easy quantity to measure for excited states,  we  consider a different approach.  Electrons tunneling between the tip and the substrate of the set up illustrated in Fig.\ref{fig1}a   can excite states of energy $\Delta$  (provided that the bias energy  $eV$ is larger than $\Delta$) when they go  through one of the magnetic atoms. The opening of a new (inelastic) tunneling channel results in an stepwise increase of the conductance $dI/dV$.  The width of these steps is proportional to the temperature $T$ because of the thermal smearing of the Fermi surfaces of the tip and the substrate.  Spin assisted tunneling 
 arising from cotunneling exchange,  makes IETS sensitive to spin excitations of the chain~\cite{Gauyacq_Lorente_psc_2012}.  The inelastic tunneling current when the tip is placed on top of atom $n$ reads~\cite{Rossier_prl_2009} 
\beqa
\hspace{-0.3cm}
I(n)={\cal T}  \sum_{m}P_m \! \sum_{m',a} \left|   \langle m| S^a_n |m'\rangle   \right|^2  
i(\Delta_{m,m'},eV),
 \label{current}
\eeqa
where $a=\{x,y,z\}$ and ${\cal T}$ is a dimensionless constant that scales linearly with the tip-$n$-adatom and adatom substrate coupling.  
$P_m$ is the 
occupation of the $|m\rangle$ state and $\Delta_{m,m'}= E_m-E_{m'}$.
$i(\Delta_{m,m'},eV)=(G_0/e)\Big[{\cal G}(\Delta_{m,m'}+eV) - {\cal G}(\Delta_{m,m'}-eV)\Big]$ is the current of a single inelastic chanel, where $G_0$ is the quantum of conductance and ${\cal G}(\omega)\equiv \omega\left(1-e^{-\beta \omega}\right)^{-1} $ is  the phase space factor ($\beta=1/k_B T$).
Finally,
the  matrix elements  of the spin operators of atom $n$,
$
\langle m| S^a_n|m'\rangle,
$
relate the current characteristics to properties of the quantum spin eigenstates.  For small current flow, the
occupations, $P_m$, are given by their thermal equilibrium values~\cite{Delgado_Rossier_prb_2010,Loth_Bergmann_natphys_2010}.
Because only the ground state is significantly occupied for $k_B T \ll \Delta_{1,0}$, the inelastic 
current provides information about  excitation energies and  matrix elements, $\langle 0| S^a_n|m'\rangle$, connecting the ground state
and certain excited states.  The form factors that control the intensity   of the step in the $dI/dV$ curve for energy $\Delta_{1,0}$  are
 $\sum_{m',a} \left|   \langle 0| S^a_n |m'\rangle   \right|^2$, where $m'$ runs over the  first excited $S=1$ states, are
shown in Fig.2(b) for different sites $n$  and  chain lengths $N$. These matrix elements are enhanced  near the edges  and the edge/center ratio increases with $N$. The asymptotic behavior (large $N$) of this  ratio should be an exponential increase in $N$, $\sim e^{N/\xi'}$, where $\xi'$ is of the order of $\xi$.  The implication of this behavior in the transport is illustrated in Fig.~\ref{fig2}(c), which shows the calculated $dI/dV$ for the different atoms in a chain of $N=16$ spins. All of them have a step at $eV=\Delta_{1,0}$, but the intensity of the step is 3 times larger at the edge than at the center.  In addition,  by measuring $\Delta_{1,0}$ as a function of $N$, it is possible to extract the spin correlation length $\xi$~\cite{White_Huse_prb_1993} from a fit with Eq.~(\ref{xi}).  The same type of spectroscopy  performed on a ring of  equidistant spins would also allow to measure of the value of the Haldane gap.

%

%

%
%

We will now consider the effects of  the single-ion anisotropy terms.  Previous theory work has addressed this issue in the context of bulk-probe  experiments~\cite{Batista_Hallberg_prb_1998,Yang_Yang_prl_2008}. Here we  tackle  the effects of these terms on the STM-IETS probe. Experiments for transition metals on Cu$_2$N show that the exchange constant is much stronger than the single-ion anisotropy terms for Mn spin chains~\cite{Hirjibehedin_Lutz_Science_2006},  but not for Fe chains where the two energy scales are similar~\cite{Loth_Baumann_science_2012}.   The difference between these two cases is that the orbital moment is quenched  for the case of Mn atoms ($S=5/2$), but it is not for the case of Fe atoms.

\begin{figure}
\includegraphics[width=0.9\linewidth,angle=0]{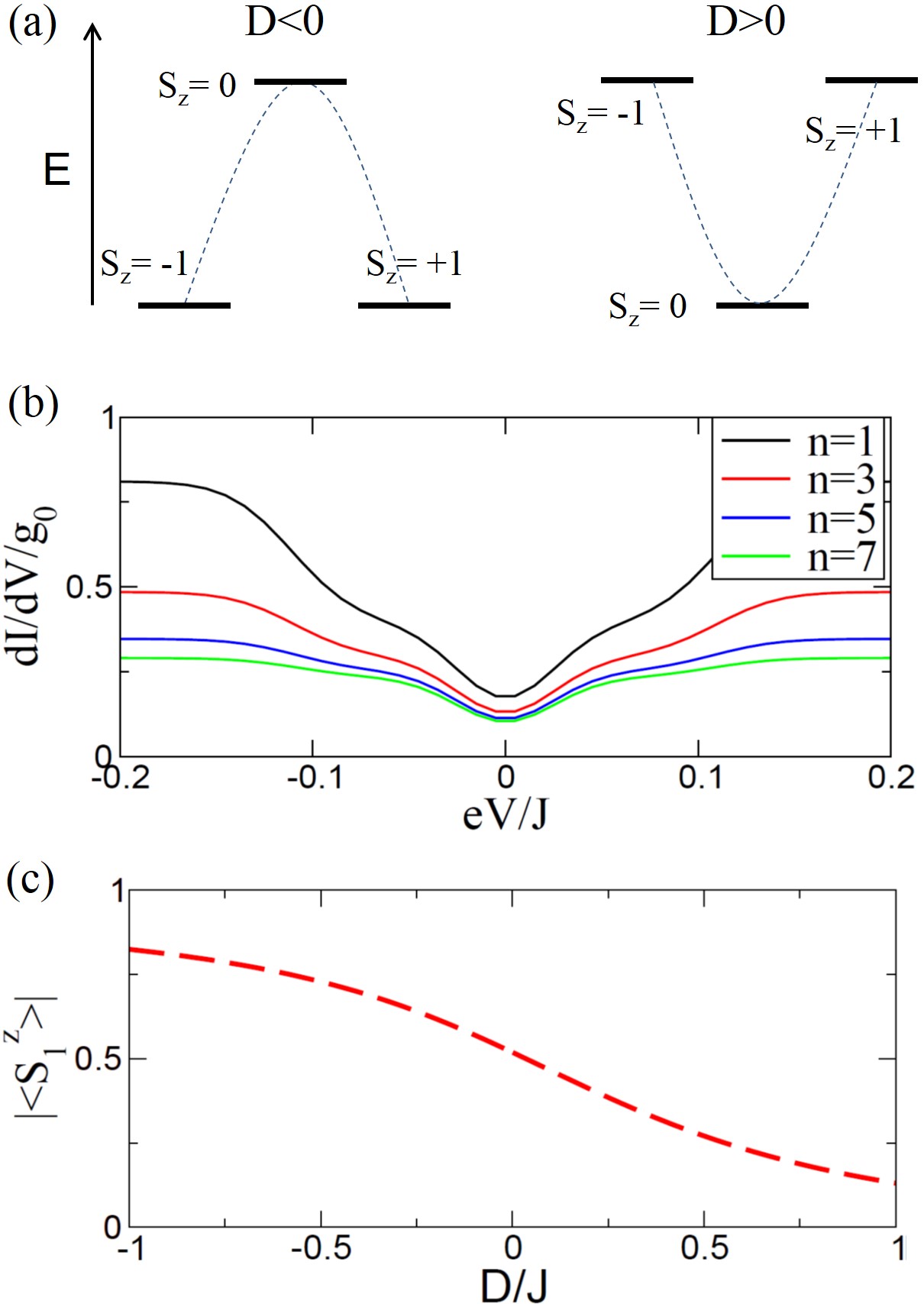}
\caption{ (Color online)\label{fig3} (a) Scheme of the energy level diagram of each spin in the chain for $D<0$ (left panel) and $D>0$ (right panel) when $J/|D|\to 0$ and $E=0$. (b) Low bias  $dI/dV$ along different positions $i$ of a $N=16$ spin chain with anisotropy parameters $D=-0.2J,\;E=0.05J$ and $T=0.01J$. (c) Expectation value $|\langle 1 |  S^z_1 | 1 \rangle |$ (red line) versus $D/J$. }
\end{figure}

As we show in Fig.~\ref{fig3}(a), the uniaxial single-ion anisotropy, $D$, splits the $S=1$ triplet into a $S^z=\pm 1$ doublet and an $S^z=0$  singlet.
Depending on the sign of $D$, either the singlet or the doublet are pushed down in energy.  The in-plane anisotropy breaks the degeneracy of the $S^z=\pm 1$ doublet. The splitting, $\Delta_{1,0}\sim E$, between the resulting eigenstates, $(| S^z=1 \rangle \pm | S^z=-1 \rangle)/\sqrt{2}$, 
can be observed with  IETS as a fine structure in the  inelastic step associated with transitions between the ground and the first excited states, provided $\Delta_{1,0}\gtrsim 5.4k_BT$, see Fig.~\ref{fig3}(b).

The local character of the edge excitations survives under the presence of small anisotropy terms $D$ and $E$. This is clear from the rapid decrease in the intensity of the  IETS signal as the tip moves from the edges  towards the center of the chain [see Fig.~\ref{fig3}(b)].  
 For large enough and negative $D$, the model undergoes a quantum phase transition into an Ising-like AFM  ground state that does not have $S=1/2$ edge states. 
Therefore, it is interesting to determine how the nature of the  edge states changes  as a function of increasing $|D|$.   For that matter, we compute the local expectation value of
the edge spin in  the lowest energy triplet state, $\langle 1 |  S^z_1 | 1 \rangle$, which should approach $1/2$ if there are fractionalized edge states. As it is shown  in Fig.~\ref{fig3}c, $|\langle 1 |  S^z_1 | 1 \rangle|$ decreases and tends to zero for positive values of $D$, while it increases and tends to $1$ for negative values of $D$. This behavior is consistent with the evolution towards a quantum paramagnet for $D> D_{c1} > 0$ and  an Ising N\'eel antiferromagnet for $D< D_{c2}<0$.


%
%

We now discuss promising materials to fabricate $S=1$ spin chains  under control and  verify our predictions. 
So far,    STM-IETS has revealed two different systems with $S=1$ moments:  Fe adatoms on InSb(110)~\cite{Khajetoorians_Chilian_Nature_2010} and Fe Phthalocyanine  (Pc) molecules on oxidized Cu.  However, the strong coupling of laterally assembled Pc molecules seems unlikely and the manipulation of magnetic atoms on semiconducting surfaces remains to be demonstrated.  The fabrication of spin chains with  Mn, Co and Fe atoms has been reported for the case of Cu$_2$N\cite{Hirjibehedin_Lutz_Science_2006,Loth_Baumann_science_2012,Otte_PhD_2008}, but these atoms have $S\neq 1$ moments.  Given that first row of  transition metals have a $+2$ oxidation state on this surface, we expect that 
Nickel or Vanadium atoms on Cu$_2$N should be good experimental candidates to test the properties of $S=1$ spin chains by STM-IETS. Naturally,  progress in this field will enlarge the number of surfaces and chemical species that can be used to engineer quantum spin chains and explore quantum magnetism at the nanoscale. A particularly interesting possibility could arise from the use of modified AFM molecular wheels~\cite{Furrer_Waldmann_rmp_2013} deposited on surfaces, which can be chemically modified to test the properties of open and closed chains.

In summary,  we are proposing a way to probe fractionalized $S=1/2$ edge states  of an individual $S=1$ spin chain, as well as to  map the spectral weight of the edge states along the chain. The proposal relies on the spectroscopical capabilities of STM that allow measuring spin excitations with atomic spatial resolution. We have shown that this technique opens  a direct access to the spectral properties of individual spin chains, which are finger prints of exotic states of matter. This simple example illustrates the potential for using this experimental technique  as a controllable artificial lab for testing fundamental properties of highly correlated systems.

 This work has been financially supported by MEC-Spain (Grant Nos. FIS2010-21883-C02-01, FIS2009-08744,  and CONSOLIDER CSD2007-0010), European Union as well as Generalitat Valenciana, grant Prometeo 2012-11. This work was carried out under the auspices of the NNSA of the US DoE at LANL under Contract No. DE-AC52-06NA25396, and was supported by the US Department of Energy, Office of Basic Energy Sciences, Division of Materials Sciences and Engineering.



\end{document}